\documentclass[]{spie}
\usepackage[utf8]{inputenc}
\usepackage[frenchb]{babel}
\usepackage[pdftex]{graphicx}
\usepackage{bmpsize}
\DeclareGraphicsExtensions{.jpg}
\usepackage{amssymb}
\usepackage{float}
\usepackage[squaren,Gray]{SIunits}
\usepackage{amsmath}

\usepackage{color}
\definecolor{blue}{rgb}{0.1,0.1,0.6}
\definecolor{orange}{rgb}{0.74,.35,0.099}
\definecolor{pale}{rgb}{0.90,0.90,0.95}
\definecolor{red}{rgb}{1.0,0.0,0.0}

\title{Gemini Planet Imager Observational Calibrations VI: Photometric and Spectroscopic Calibration for the Integral Field Spectrograph}

\author{J\'er\^ome Maire\supit{a}, Patrick J. Ingraham\supit{b}, Robert J. De Rosa\supit{c,d}, Marshall D. Perrin\supit{e}, Abhijith Rajan\supit{c},
Dmitry Savransky\supit{o},   Jason J. Wang\supit{g}, Jean-Baptiste
Ruffio\supit{h}, Schuyler G. Wolff\supit{i},
  Jeffrey K. Chilcote\supit{j}, Ren\'e Doyon\supit{k}, James R. Graham\supit{g},  Alexandra Z. Greenbaum\supit{i}, Quinn M. Konopacky\supit{a}, James E. Larkin\supit{j}, Bruce A. Macintosh\supit{b,f},
Christian Marois\supit{l}, Max Millar-Blanchaer\supit{m}, Jennifer
Patience\supit{c}, Laurent A. Pueyo\supit{e},  Anand
Sivaramakrishnan\supit{e}, Sandrine J. Thomas\supit{n},  Jason L.
Weiss\supit{j} \skiplinehalf
\supit{a} Dunlap Institute for Astronomy \& Astrophysics, University of Toronto, 50 St. George Str. Toronto, ON M5S 3H4, Canada; \\
\supit{b} Stanford University, Kavli Institute for Particle Astrophysics and Cosmology, Stanford, CA 94305, USA; \\
\supit{c} School of Earth and Space Exploration, Arizona State University, PO Box 871404, Tempe, AZ 85287, USA; \\
\supit{d} School of Physics, College of Engineering, Mathematics and Physical Sciences, University of Exeter, Stocker Road, Exeter, EX4 4QL, UK; \\
\supit{e} Space Telescope Science Institute, 3700 San Martin Dr, Baltimore, MD 21218, USA \\
\supit{f} Lawrence Livermore National Lab, 7000 East Ave., Livermore, CA 94551, USA; \\
\supit{g} Department of Astronomy, UC Berkeley, Berkeley CA, 94720, USA \\
\supit{h} SETI Institute, 189 Bernardo Avenue, Mountain View, CA 94043, USA \\
\supit{i} Johns Hopkins University, Physics \& Astronomy Department, Baltimore, MD, 21218, USA; \\
\supit{j} Division of Astronomy, UCLA, Los Angeles, CA 90095, USA; \\
\supit{k} Universit\'e de Montr\'eal, Montr\'eal QC H3C 3J7, Canada; \\
\supit{l} National Research Council of Canada Herzberg, 5071 West
Saanich Road,
Victoria, BC V9E 2E7, Canada; \\
\supit{m} University of Toronto, Toronto, 50 St. George Str. Toronto, ON M5S 3H4, Canada; \\
\supit{n} NASA Ames Research Center, Mountain View, CA 94035, USA; \\
\supit{o} Sibley School of Mechanical and Aerospace Engineering, Cornell University, Ithaca NY USA 14853.\\
}

\authorinfo{Further author information: (Send correspondence to J.M.)\\J.M.: E-mail: maire@dunlap.utoronto.ca\\  P.J.I.: E-mail:patricki@stanford.edu}

\begin{document}
  \maketitle

%%%%%%%%%%%%%%%%%%%%%%%%%%%%%%%%%%%%%%%%%%%%%%%%%%%%%%%%%%%%%
\begin{abstract}
% \todo{MP: this abstract is 60\% generic text about GPI, only 40\% about what's in this specific paper. Should be revised to be more specific and give some of the key conclusions about GPI performance. JM: changed}
The Gemini Planet Imager (GPI) is a new facility instrument for the
Gemini Observatory designed to provide direct detection and
characterization of planets and debris disks around stars in the
solar neighborhood. In addition to its extreme adaptive optics and
coronagraphic systems which give access to high angular resolution
and high-contrast imaging capabilities, GPI contains an integral
field spectrograph providing low resolution spectroscopy across five
bands between 0.95 and 2.5 $\mu$m. This paper describes the sequence
of processing steps required for the spectro-photometric calibration
of GPI science data, and the necessary calibration files. Based on
calibration observations of the white dwarf HD 8049~B we estimate
that the systematic error in spectra extracted from GPI observations
is less than 5\%. The flux ratio of the occulted star and fiducial
satellite spots within coronagraphic GPI observations, required to
estimate the magnitude difference between a target and any resolved
companions, was measured in the $H$-band to be $\Delta m =
9.23\pm0.06$ in laboratory measurements and $\Delta m = 9.39\pm
0.11$ using on-sky observations. Laboratory measurements for the
$Y$, $J$, $K1$ and $K2$ filters are also presented. The total
throughput of GPI, Gemini South and the atmosphere of the Earth was
also measured in each photometric passband, with a typical
throughput in $H$-band of 18\% in the non-coronagraphic mode, with
some variation observed over the six-month period for which
observations were available.
%Ongoing development of the extraction algorithm.
We also report ongoing development and improvement of the data cube
extraction algorithm.
%This paper describes the sequence of processing steps required for the spectro-photometric calibration of GPI science data, and the necessary calibration files. Absolute calibration uncertainties and stability are investigated by means of spectro-photometric standard star observations.

\end{abstract}

%>>>> Include a list of keywords after the abstract


%%%%%%%%%%%%%%%%%%%%%%%%%%%%%%%%%%%%%%%%%%%%%%%%%%%%%%%%%%%%%
\section{Introduction} \label{sec:intro}

%  The detection and characterization of planets outside the solar system have become one of the top-level priorities for research in astronomy since the first indirect detections of planets two decades ago. Hundreds of planetary systems have been discovered around other stars, mainly from Doppler shifts of spectral features from stars (Mayor et al. 1995\cite{mayor95}) and from transits of planets across the face of their parent stars (Charbonneau 2000\cite{charbonneau00}). High-contrast direct imaging and spectroscopy complements our understanding of the formation and evolution processes of planetary systems by exploring and revealing the characteristics of exoplanet population at large distances (beyond several AU) from their parent stars.
% \todo{RDR: microlens or micro-lens? similarly for micro-spectra. we should be consistent through the paper, and all the other papers i guess. JM:OK, dunno which one is best but I replaced every microlens with micro-lens, so it is consistent}
The combination of high angular resolution and high-contrast
imaging, made possible by tremendous improvements in the performance
of adaptive optics (AO) systems \cite{Macintosh06, Poyneer07,
poyneerthis} and developments of coronagraphic
capabilities\cite{Guyon2007, Sivaramakrishnan2010, Soummer2007}, is
opening a new window into studying the formation and evolution
processes of planetary systems by exploring and characterizing the
exoplanet population at large orbital separation (beyond several
AU). Direct imaging observations are particularly valuable because
they enable spectroscopic characterization of the detected
planetary-mass
companions\cite{Konopacky2013,Oppenheimer2013,Bonnefoy2014}. Several
exoplanets have been already directly imaged \cite{ Marois08,
Kalas08,  Lagrange10, Lafreniere10, Marois10}, but the required
extreme contrast ratio required at very small angular separations
remains a major obstacle in the detection of new companions.
% exoplanet structure
% through the composition and temperature of  planetary atmospheres.

  The Gemini Planet Imager\cite{Macintosh08, Graham07, Macintosh12, macintosh2014first} (GPI) is a new high angular resolution extreme adaptive optics instrument that provides unprecedented high-contrast capabilities.  After a successful installation at Gemini South in October 2013, the GPI team performed four commissioning observing runs to assess the performance of the instrument, followed by an early science observing run open to the community. GPI's science instrument is an integral field spectrograph (IFS) that can observe in one of the near-infrared $Y$, $J$, $H$, $K1$, $K2$ filter bandpasses. In its spectral mode, the IFS makes use of a lenslet array  and a prism \cite{Larkinthis,Chilcote12} to subdivide its 2.7''$\times$2.7'' field of view into 192$\times$192 spectra, with a plate scale of 14.3 mas/pixel \cite{konopackythis}, and a spectral resolving power ranging from $\lambda/d\lambda=35$ to 90\cite{Chilcote2012}, depending on the band. Because of the increased resolution at longer wavelengths the $K$ band is split into two partially overlapping bandpasses for GPI; $K1$ ($1.9-2.2$~$\mu$m) and $K2$ ($2.1-2.4$~$\mu$m). The micro-lens array is rotated by 24.8 degrees with respect to the detector to maximize the number of spectra on the detector, and to optimize the distance between them. In polarimetric mode, the spectral prism is replaced by a Wollaston prism and a rotating half-wave plate is inserted into the beam in order to measure the polarization of the incoming light\cite{Perrin10}.

  %\todo{Cite Perrin et al. 2010 SPIE for design of the polarimeter. Sloan's paper has no hardware details.}\cite{Wiktorowiczthis}.
  % FYI - SLoan's paper is only about instrumental polarizatoin calibration. Reduction is discussed in my ApJ paper to be submitted.
  % Totally fine to cite that one if you want, but all this background detail is not essential here to cite everything.
  % Will leave out reduction of polarimitry data since reduction isn't really the focus of this paper.
 The reduction of GPI IFS raw science images, containing $\sim$37,000 spectra, is a complex process that involves the need for accurate calibrations. The GPI Data Reduction Pipeline\cite{Maire2010, perrinthis} (DRP) has been developed to produce calibrated datacubes with minimal processing artifacts, ready for scientific analysis. This paper describes the steps necessary to perform photometric (Section \ref{sec:cal}) and spectroscopic (Section \ref{sec:spec}) calibration of reduced GPI IFS data. The absolute calibration of the fiducial satellite spots, necessary for measuring companion magnitudes and contrast, is discussed in Section \ref{sec:spot}. The throughput of the instrument is measured in Section \ref{sec:throughput}, and a new method of spectral extraction implemented in the pipeline utilizing micro-lens PSFs to reduce systematics is presented in Section \ref{sec:ongoing}.
 %disentangle the flux variations along each spectrum? (should this be mitigate cross-talk).

%   Investigations of the brightness of distant exoplanets are necessary to understand their structure and behavior.
%  The photometric calibration steps aims to measure the amount of light energy collected by the telescope and GPI instrument from a specific target. Detailed studies of the time variations in brightness of exoplanets and brown dwarfs give insights into their atmosphere structure and composition.

%   Polarimatric reduction process is detailed in Wiktorowicz et al. 2014\cite{Wiktorowiczthis}. Speckle suprresion algorithms are discussed in Marois et al. 2014\cite{Maroisthis}.
%   These methods will continue to evolve as GPI operations further mature over years to come.

\section{GPI Photometric calibration} \label{sec:cal}
% Remind few relevant things about Gemini-GPI instrument, IFS and data (maybe +Fig: raw data \& extracted slice to give a better idea)

% [datacube extraction]

% important to introduce the sat spots here - and how they are used etc.
%     - recipes used in pipeline
%     - units
%     - few things about sat spot, wavelength solution (need to be good!)
%     (+Fig: maybe a diagram showing all the calibration steps)

%     SATURATION table
Several steps are necessary to transform the raw two-dimensional GPI
IFS detector images containing the micro-spectra into
three-dimensional spectrophotometrically calibrated datacubes. The
first steps are for mitigating detector noise that affects the raw
images to remove correlated noise and detector artifacts. A detailed
characterisation of this source of noise and methods to reduce its
effect are fully described in Ingraham et al.
2014\cite{Ingraham1this}. These corrections are more naturally
performed directly on the IFS raw two-dimensional images, since this
noise is directly engendered by, or related to the detector.

% \todo{This entire paragraph is redundant with the detectors paper. I don't see why its necessary to repeat this information here? Why not just say "Detector systematics are removed by the GPI pipeline using the methods described in [reference]" and leave it at that? JM: true, but we can not ask the reader to read all the papers. Otherwise your comments suggest that we should remove the general description of GPI since it is on other papers as well. I think this part on detector systematics is important to remind since it may affect the photometry.}
HAWAII-2RG detectors generate dark current, smaller than
0.1~e$^{-}$/s/pixel, that produces a mean fixed pattern in the image
during an exposure. Master dark frames are produced daily, at
exposure times ranging from 1.45~s to 120~s, which are used to
correct observations with the ``Subtract dark background" primitive
within the GPI DRP. Other sources of noise, such as high-frequency
striping due to readout electron noise and microphonic vibrations
induced by the cryo-coolers\cite{Chilcote2012} are removed using the
``Destripe science image" primitive on images that contain enough
non-illuminated pixels to derive a noise model. Defective pixels are
identified using the periodically updated GPI bad pixel map, and
their intensities are corrected by the primitive ``Interpolate bad
pixels in 2D frame" using surrounding pixel intensities to
interpolate the intensity of the defective pixel. Other effects,
such as persistence can also be corrected\cite{Ingraham1this},
whereas effects such as non-linearity are best avoided by
maintaining the total number of counts per pixel below $\sim$17,000
ADU/coadd, or by allowing the persistence to diminish after a
sequence of saturated exposures.

The datacube extraction procedure requires the location of the
$\sim$37,000 spectra within the raw image, for each filter used, to
an accuracy of better than a tenth of a pixel\cite{Schuylerthis}.
The separation between the individual spectra on the raw image is
$4.5$ pixels at the center of the image. Field-dependent aberrations
induced by the re-imaging optics after the lenslet array result in
the tilt of the micro-spectra changing as a function of their
position on the detector. This results in the inter-spectra
separation varying slightly over the field. Measured wavelength
solutions for each band are provided from the reduction of Gemini
GCAL xenon and argon
% \todo{RDR: pretty sure elements are lower-case: JM any way is fine with me}
lamps. The locations of the spectra given by the wavelength solution
are corrected for instrumental flexure effects that shift the
spectra on the detector as the orientation of the instrument
varies\cite{Schuylerthis}.
% \todo{RDR: this following sentence is a bit disjointed from the previous one, it could go first, and then the aperture could be described.JM: Thanks ! done}
Light passing through each micro-lens is diffracted and the
resulting point spread function (PSF) spreads over several detector
pixels, with a typical 1.5 pixel full width at half maximum (FWHM)
in $H$-band ($1.5-1.8$~$\mu$m). The original method of datacube
extraction implemented into the GPI pipeline proceeds by integrating
the signal over a rectangular 1$\times$3 pixel aperture centered on
the spectrum, a process which is repeated for each pixel along the
dispersion axis to produce the final spectrum. The length of the
photometric rectangular aperture was therefore chosen to integrate
the maximal number of pixels covering the extent of the micro-lens
PSF at the wavelength of interest, and small enough to limit the
contamination from the PSF wings coming from surrounding spectra.
Once the spectra have been separately extracted, they are
interpolated onto a common wavelength axis and assembled into a
datacube. The steps discussed above are accomplished using a
combination of primitives ``Load Wavelength Solution", ``Update Spot
Shifts for Flexure", ``Assemble Spectral Datacube" and ``Interpolate
Wavelength Axis", and allows for the pipeline to produce a reduced
datacube in few seconds using a standard desktop or laptop computer.
Since no matrix inversion is required, the method is extremely
robust, avoiding any noise amplification or convergence issues that
can happen with standard inversion methods. However, this method is
partially affected by the small spacing between spectra, especially
at the corners of the field-of-view, where the spectra are highly
tilted with respect to the spectral axis and contamination from
surrounding spectra becomes more significant. Based on simulations
of raw IFS images where the light coming from one micro-lens was
artificially blocked, we measured the contamination from adjacent
micro-lenses to be less than 2\%.
% \todo{Can we say anything quantitative about how much this contamination is? RDR: Also we should mention that it becomes worse at longer wavelengths.JM: I mention simulation. That's true that it becomes worse at longer wavelength but the longest wavelength on a spectrum can affect spectral channels at shorter wavelengths on the neighboor spectrum, due to the configuration of the spectra onto the dectector}

  \begin{figure}[h!]
  \begin{center}
        \includegraphics[width=11cm,natwidth=530,natheight=357]{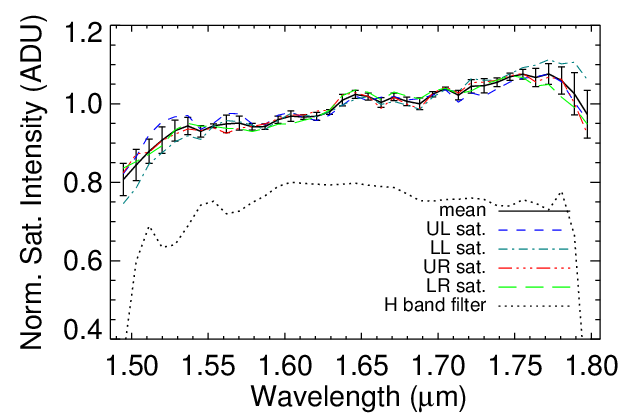}\\
  \end{center}
  \caption{ \label{fig:wd_sats} The individual extracted spectra for the four satellite spots, normalized to have equal integrated intensity over the band. The uncertainties from satellite spectra are propagated into the extracted spectra of companions. The UL, UR, LL, LR acronyms indicate the upper left, upper right, lower left, and lower right satellite, respectively. The dotted curve represents the H-band filter profile with the peak transmission normalized to 0.8 for ease of comparison to the satellite spot spectra.
%   \todo{JRG asks if we could plot the filter band pass on the same axes for comparison}
  }
  \end{figure}

As the micro-lens PSF extends over several pixels along the spectral
axis as well, internal cross-talk between spectral channels of the
same micro-spectrum occurs when using the integration method,
especially for spectra with strong features. This cross-talk impacts
adjacent spectral channels and affects the spectral resolution. The
length of the spectra, about $\sim$18 pixels in the $H$-band,
% to $\sim$20 pixels in the $K2$-band,
results in a detector-pixel-limited Nyquist sampled spectral
resolution about $\sim$$50 \pm 6$ in $H$-band,
% \todo{RDR: values for Y and K2 are given, then values for H are given, we should give the resolution in Y and K2 as they represent the limiting cases. JM:okay I left only H-band}
depending the location in the field-of-view. The datacube generated
by the GPI DRP contains 37 spectral channels which are spectrally
oversampled by a factor $\sim$2 relative to the detector sampling.
% 4, if one considers the FWHM of the micro-lens PSF as a resolution criterion.
It is possible to reach slightly better resolving power by taking
into account of the micro-lens PSF extent in the datacube extraction
process (see Section \ref{sec:ongoing}).

It should be noted that the flat-field correction methods for GPI
data are still in development. Multiple factors contribute different
types of flat fields, including in particular the lenslet array flat
and the detector pixel flat. Because of shifts of the spectra due to
flexure effects, the detector pixels illuminated when observing a
daytime calibration flat-field are not necessarily the same ones
illuminated in any particular science image of interest. This
results in different pixel sensitivities in the flat-field and
science frames. Consequently, it is recommended to only use a master
flat-field that has exactly the same locations of spectra in the raw
frames than those for the science image to be corrected. This could
be obtained by taking images using the flat-field lamp just before
and/or after the target observations at similar elevations, under
the assumption that the flexure effect will remain stable during the
course of those observations.

Separately, a lenslet flat field can be derived to account for the
varying illumination through the lenslet array. A handful of
scattered lenslets have lower than average throughput which can be
compensated for.
%\todo{BM - how many roughly}.
Furthermore, an artifact from the manufacturing process for the
lenslet array is that every 30th row and 50th column has slightly
higher throughput than average, each accompanied by a slightly
fainter adjacent row or column. We interpret these as the boundaries
between replicated blocks of 30$\times$50 lenslets. The observed
effect is small, with a throughput variations for the affected rows
and columns of $\sim0.5\%$. This can be compensated for using the
lenslet flat field, though it is a small enough correction to be
negligible in many circumstances.

In coronagraphic mode, the absolute photometric calibration takes
advantage of the presence of fiducial satellite spots, produced by
the diffraction of light from a square grid printed onto the
apodizer\cite{Marois06,Sivaramakrishnan06}. These four satellite
spots represent four images of the unocculted host star, scaled in
intensity by a known factor, or grid ratio, measured in the lab.
Together with the spectral type and apparent magnitude, they permit
an astrometric and photometric calibration of the data. The
laboratory and on-sky measurement of the grid ratio is given in
Section \ref{sec:spot}, with a detailed discussion on the stability
of the photometry of these fiducial spots given in Wang et al. 2014
(this proceedings\cite{Wangthis}). Once the fiducial spot mean
intensity has been measured in each slice of the datacube, it is
possible to use the grid ratio to estimate the number of counts that
an image of the unocculted star would have produced.

%\begin{table}[h]
% \begin{center}
% \caption{The physical units of calibrated datacubes produced by the pipeline.}
% \label{tab:units}
% \begin{tabular}{|c|}
% \hline
%   {\bf DRP available units}   \\
%      \hline
%     ph/s/nm/m$^2$\\
%     \hline
%     Jy\\
%     \hline
%     W/m$^2$/$\mu$m \\
%     \hline
%     ergs/s/cm$^2$/\angstrom \\
%     \hline
%     ergs/s/cm$^2$/Hz \\
%      \hline
%\end{tabular}
%\end{center}
%\end{table}

The DRP primitive ``Calibrate Photometric Flux" uses the grid ratio,
photometry of the spots at each wavelength, gain of the detector,
exposure time, collecting area of the telescope, and the magnitude
and spectral type of the occulted star to perform the photometric
calibration, compensating for the instrumental and atmospheric
transmission functions and converting measured ADU into a variety of
commonly-used physical units (photons/s/nm/m$^2$, Jy,
W/m$^2$/$\mu$m, ergs/s/cm$^2$/$\angstrom$ and ergs/s/cm$^2$/Hz). The
pipeline contains a copy of the Pickles stellar spectral flux
library \cite{Pickles1998} for a wide range of stellar spectral
types to be used as part of the calibration, but also permits the
use of user-defined spectra. Surface brightness units may be
obtained simply by normalizing by the area of each lenslet, 14.3
mas$^2$.
% In the future, we plan to expand the pipeline to also support units of surface brightness units in addition to the flux units.
% \todo{Can't we just say "Surface brightness units may be obtained simply by normalizing by the area of each lenslet, (14.3 mas)$^2$". JM:done}

\section{GPI Spectroscopic Calibration} \label{sec:spec}

Determining the accuracy of the spectroscopic extraction of GPI
spectra is required to assign uncertainties to our spectral data.
The primary challenge in characterizing the spectral extraction
performance is to understand the systematic errors that may result
from each step in the reduction process. Observations of standard
stars are used to test the quality of spectral extraction, but GPI
is observing in a parameter space that is beyond the separation and
contrast limits of previous imagers. Therefore only a small number
of previously observed standards exist; several of which are cooler
L and T-type brown dwarfs that do not have well-understood spectra.
Of the known close companions, the few white dwarfs are most
attractive for spectral characterization given their well-understood
atmospheres.

As part of the GPI Verification and Commissioning stage, the white
dwarf companion to HD 8049 was observed to asses the accuracy of
spectroscopic retrieval from GPI IFS data reduced using the DRP.
Previous observations\cite{Zurlo2013} have constrained the white
dwarf (HD 8049 B) to have an effective temperature of
$18,800\pm2,100$~K, and the spectrum is consistent with a blackbody
of the same temperature. This target represents an ideal calibration
source for  GPI\footnote{Another potential white dwarf companion to
HD~114174 \cite{Crepp13a} has also been observed, but its
temperature is not yet well constrained.}. The dataset consists of
five 90s $H$-band exposures of HD~8049 taken on December 10, 2013.
For this data, zero-point offsets to the wavelength solutions were
determined using an argon arclamp image taken immediately following
the exposures. The dark subtracted detector images had their
micro-spectra extracted using the 3-pixel box extraction algorithm.
The individual cubes were then stacked in two different ways. The
first method was a straight median stack resulting in a higher
signal to noise of the satellites and speckles, but a spectral
smearing of the companion. The second stack was created by first
rotating each image such that North was in the vertical direction
prior to performing the median. This ensured a higher
signal-to-noise ratio (SNR) of the companion, while blurring the
speckles and satellites. These two cubes were then input into the
DRP primitive ``Calibrate Photometric Flux" that both detects the
satellites and extracts their spectra, then in combination with a
designated spectral type and apparent magnitude of the primary star,
outputs a spectrophotometrically calibrated version of the derotated
median cube. Figure \ref{fig:wd_sats} shows the four extracted
spectra of the satellite spots normalized to have the same
integrated intensity, with the mean value plotted as the solid black
line and the error bars signifying the standard deviation between
the four points. In the case of Figure \ref{fig:wd_sats}, the
spectra have been normalized for ease of comparison over the full
wavelength range. The error bars shown here do not include the error
component of each individual satellite which is derived from looking
at the standard deviation of an arc shaped region surrounding the
satellite, convolved by the extraction aperture.

\begin{figure}[h]
  \begin{center}
    \includegraphics[width=9cm,natwidth=530,natheight=357]{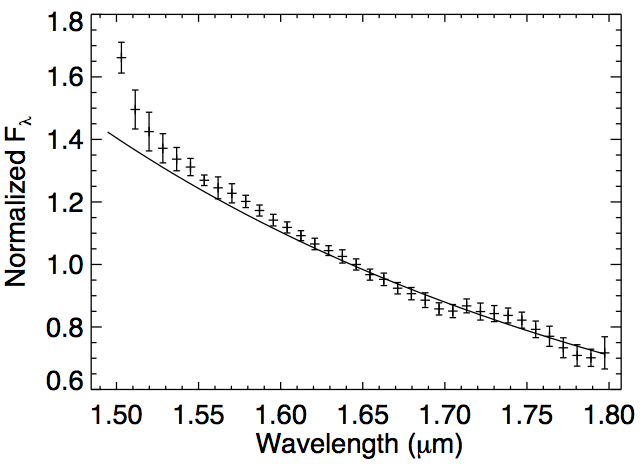} \\
  \end{center}
  \caption{\label{fig:wd1} The calibrated $H$-band spectrum of HD 8049 B overplotted with a 18,800 K blackbody (solid line).}
\end{figure}
The GPI pipeline extracts the spectral data over the range in
wavelengths that has a filter transmission of 10\% or higher. This
results in a significant decrease in the SNR towards the edge of the
bands, where the filter transmission drops sharply. Large systematic
errors may result when calibrating the cube towards the filter edges
in the regions of dropping transmission, particularly if there are
small errors in the wavelength calibration. For these reasons, we
advise users to exercise caution and trim their data according to
the quality and SNR of their spectra and wavelength calibration. We
have opted to leave the wavelengths with 10\% transmission as part
of the cubes intentionally so that users may exercise their own
judgement on the usable wavelength region, rather than imposing
sharp cutoff by the pipeline at some higher throughput.

Obtaining an absolute flux calibration requires knowledge of the satellite spot to star flux ratio, a value which has shown variability in the past in shape and is the subject of investigation as part of GPI commissioning. Further details regarding the satellite spots are given in Section \ref{sec:spot} and in Wang et al. 2014 (this proceedings\cite{Wangthis}). The pipeline is currently using satellite spot to star flux ratios that were measured during the integration and testing of the instrument, although on-sky measurements obtained during commissioning show similar values (shown in Table \ref{tab:grid_ratio} and discussed in section \ref{sec:spot}). The pipeline values will be updated to reflect new on-sky measurements in an upcoming pipeline release. %\todo{This leaves the reader hanging - "theres an error but we won't tell you any details or how much it is". Can we say something more like "The spectra produced by the pipeline may be subjet to systematic bias of a few percent" or something like that?}, and therefore we present only relative flux calibrated spectra.

Extracting a spectrum from a calibrated datacube is performed using
the DRP primitive, ``Extract 1d spectrum." This primitive uses the
same extraction aperture (by default) as is in extracting the
satellite spot spectra to ensure no introduction of systematic
errors from aperture corrections. The uncertainty computed by the
primitive combines the uncertainty from the satellite spot spectra
along with an uncertainty of the noise surrounding the companion.
Figure \ref{fig:wd1} shows the calibrated $H$-band spectrum of HD
8049 B overplotted with an 18,800 K blackbody. Overall, the fit is
of high quality, but exhibits a reduced $\chi^2$ of 2.00 due to the
data points on the blue end of the spectrum that are in a region of
low throughput and obvious victim of the systematic error discussed
above. If we restrict the wavelength range of the data to the region
where transmission function is smooth and above 50\%
($1.51~\mu\text{m}<\lambda<1.79~\mu\text{m}$), the resulting reduced
$\chi^2$ is 1.07, demonstrating that the model is well represented
by the data. Another systematic present in the data is a small
increase in flux near 1.73~$\mu$m.  This is believed to be a
systematic bias due to datacube extraction, but analysis into this
is ongoing. Current estimates of the systematic errors contained in
the pipeline, arising primarily from the extraction algorithm, are
estimated to be $<5\%$ based on this dataset. More advanced
extraction algorithms, such as what is described in section
\ref{sec:ongoing} and by Draper et al. 2014 (this proceedings
\cite{Zackthis}) are in development and are expected to
significantly decrease the systematic errors.

\section{GPI Fiducial Spot Calibration}
\label{sec:spot}
\begin{figure}[h!]
\begin{center}
\includegraphics[width=12cm]{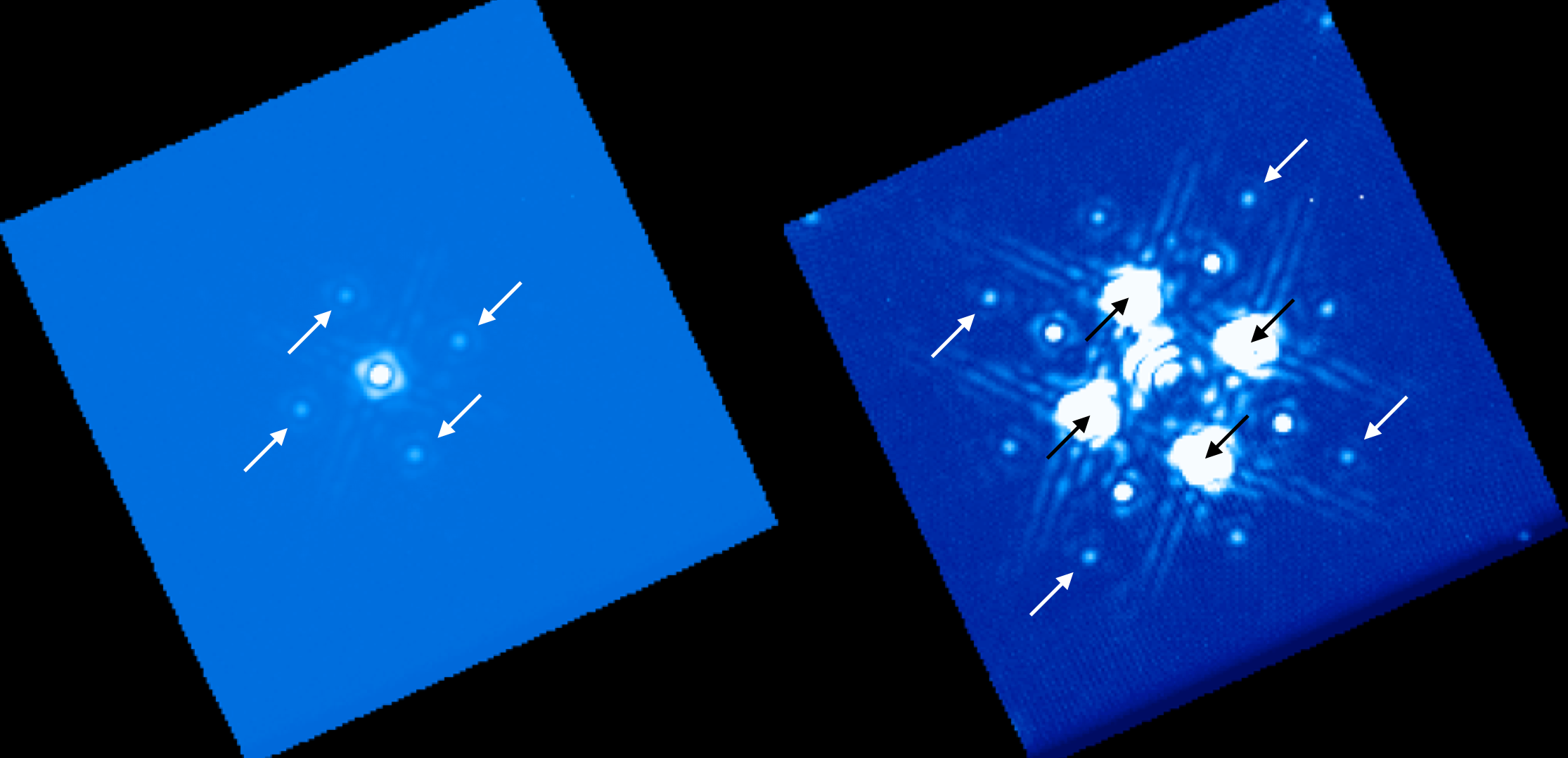}
\end{center}
\caption{\label{fig:lab_spots}{\it(left panel)} Laboratory
measurement of the unocculted PSF (image center) using a neutral
density filter to prevent saturation. As the satellite spots are too
faint to be detected, four additional reference spots (indicated)
were generated by putting vertical and horizontal cosine waves on
the deformable mirror. {\it(right panel)} Measurement of the
occulted PSF in which the satellite spots are detected (indicated by
white arrows), with an exposure time selected to prevent saturation
of the reference spots (indicated by black arrows). Due to the
removal of the neutral density filter, the reference spots are
significantly brighter. The flux ratio of the central PSF and the
satellite spots was calculated using the reference spots as a
proxy.}
\end{figure}
\begin{table}[h]
\centering \caption{Measurements of the grid ratio and corresponding
magnitude difference}
\begin{tabular}{ccccc}
\hline
 Target & Date  & Filter & Grid Ratio   & $\Delta m$ (mags)  \\
\hline \hline
Lab Measurement & 2013-02-11 & $Y$ & $4280\pm 220$ & $9.08\pm 0.05$ \\
Lab Measurement & 2013-02-11 & $J$ & $5560\pm 180$ & $9.36\pm 0.03$\\
Lab Measurement& 2013-02-11 & $H$ & $4910\pm 250$ & $9.23\pm 0.06$\\
Lab Measurement& 2013-02-11 & $K1$ & $3700\pm 240$ & $8.92\pm 0.07$\\
Lab Measurement& 2013-02-11 & $K2$ & $5200\pm 1020$ & $9.29\pm 0.21$\\
$\beta$ Pic & 2013-11-18    & $H$ & $5430 \pm 1790$ & $9.34\pm 0.36$ \\
$\beta$ Pic & 2013-12-10    & $H$ & $5400 \pm 950$  & $9.33\pm 0.19$ \\
$\beta$ Pic & 2013-12-11    & $H$ & $4570 \pm 1940$ & $9.15\pm 0.46$ \\
HD 118335   & 2014-03-25    & $H$ & $6150 \pm 960$  & $9.47\pm 0.17$ \\
\hline
\end{tabular}
\label{tab:grid_ratio}
\end{table}
To measure the magnitude difference between an occulted star and a
companion candidate, the ratio between the stellar flux and the
fiducial satellite spots needs to be accurately determined.
Laboratory measurements of the star to satellite spot ratio were
made during the integration and testing of the instrument using a
super-continuum source. As the dynamic range of the detector is not
sufficient to measure the satellite spots while keeping the
unocculted artificial star unsaturated, a two-step procedure was
used. Firstly, a series of measurements were taken using a
high-order neutral density filter, during which vertical and
horizontal cosine waves were put onto the deformable mirror,
producing a final image with an unocculted central PSF and four
relatively bright reference spots (Figure \ref{fig:lab_spots}, left
panel). Due to the limited dynamic range, the satellite spots were
not detected in these measurements. Secondly, a series of
measurements were obtained with the occulting mask and without the
neutral density filter. In these images the central PSF was
occulted, the reference spots generated due to the pattern on the
deformable mirror were not saturated, and the satellite spots were
bright enough to be detected (Figure \ref{fig:lab_spots}, right
panel). The grid ratio was then measured as the product of the ratio
of the central PSF to the reference spots within the first set of
measurements and the ratio of the reference spots to the satellite
spots within the second set of measurements. This procedure was
repeated for each filter, and the results are given in Table
\ref{tab:grid_ratio}. The stated uncertainties are the standard
deviation of each set of measurements, and do not take into
uncertainties caused by flat field effects, data cube extraction
algorithm systematics, or imperfect spot or satellite source
extraction due to PSF variations across the field-of-view.
%\todo{WHat does this really mean, imperfections in the data analysis? Are there specific factors you know that are not correct in the analysis, or is this just a handwave? Better to say here how the uncertainties were derived. I.e. the uncertainties listed in the table are the 1-sigma standard deviation of the measurements from the 4 spots, or are estimated from the variance of ten different exposures or whatever it is you actually did.}%\todo{Was the ND filter transmission actually characterized "poorly" in the sense that we know it was not a good job? You want to acknowledge the limits of your own knowledge while not coming across as saying your own work is poor}.

The ratio of the flux of satellite spots to the flux of the star
behind the occulting mask was also measured on-sky using two
different techniques. For both sets of measurements, the
observations were reduced using the standard primitives within the
GPI DRP, with the background in each image being subtracted by
applying a high-pass filter, with the satellite spots and companion
(if applicable) being masked to prevent self-subtraction. The first
technique utilized coronagraphic observations of $\beta$~Pic
obtained in November and December 2013, in which the $H$-band flux
of the planetary-mass companion $\beta$~Pic~b was measured. From
this, the unocculted flux of $\beta$~Pic~A was estimated using the
literature $H$-band magnitude difference between $\beta$~Pic~A and
b\cite{Currie2013}. The grid ratio was then calculated from the
estimated $\beta$~Pic~A flux and the average of the four satellite
spots within each wavelength slice of the reduced data cubes (Table
\ref{tab:grid_ratio}). The second technique involved obtaining
unocculted observations of HD~118335, in which the stellar flux was
kept below the saturation limit of the detector, immediately
followed by a coronagraphic sequence with a significantly higher
exposure time to ensure a high signal-to-noise ratio on the
satellite spots. From this sequence of observations, the grid ratio
can be directly measured by performing aperture photometry on the
star within the unocculted sequence, and on the satellite spots
within the coronagraphic sequence. A preliminary analysis of these
data reveals a grid ratio which is consistent with the value
measured using the $\beta$~Pic observations, but marginally higher
than the laboratory measurements (Table \ref{tab:grid_ratio}). It is
not currently known whether this discrepancy is due to a systematic
error in the analysis of the laboratory or on-sky data, or due to a
fundamental difference in the behaviour of the satellite spots
between the two sets of measurements. As additional measurements are
made, the GPI DRP will be updated to reflect the most recent
estimates of the grid ratio within each photometric band and their
uncertainties. Additional discussion regarding the stability of the
grid ratio can be found in Wang et al. 2014 (this
proceedings\cite{Wangthis}).

\section{GPI throughput}
\label{sec:throughput}

\begin{table}[h]
\centering \caption{List of calibrator targets used to measure the
telecope and instrument throughput
% \todo{Let's not have these tables right on top of each other, it doesn't look very good. JM: this have been corrected by someone.}
}
\begin{tabular}{cccccccc}
\hline
 Target & Spectral Type & Date  & Filters   & AO Status & Obs Mode & Airmass & Ref. \\
\hline \hline
HD 20619 & G2V &  2013-11-17 & $YJHK1K2$ & Open & Direct & 1.16--1.19 & \cite{Rayner2009}\\
HD 51956 & F8Ib & 2013-11-17 & $YJHK1K2$ & Open & Direct & 1.17--1.21 & \cite{Rayner2009}\\
HD 51956 & F8Ib & 2013-11-17 & $YJHK1K2$ & Open & Unblocked & 1.18--1.21 & \cite{Rayner2009}\\
HD 8049 B & WD &  2013-11-17 & $HK1K2$ & Closed & Coron & 1.05-1.15 & \cite{Zurlo2013}\\
HD 126146 & A0V & 2014-03-20 & $YJHK1$ & Closed & Direct & 1.20--1.27 & \cite{Bohlin2007}\\
HD 115617 & G7V & 2014-03-24 & $YJHK2$ & Open & Direct & 1.09--1.14 & \cite{Rayner2009}\\
HD 190285 & A0V & 2014-05-11 & $YJHK1K2$ & Closed & Direct & 1.02--1.04 & \cite{Bohlin2007}\\
HIP 68209 & A0V & 2014-05-12 & $YJHK1K2$ & Closed & Direct & 1.45--1.47 & \cite{Bohlin2007}\\
\hline
\end{tabular}
\label{tab:throughput}
\end{table}

The throughput of the GPI instrument, the Gemini telescope and the
Earth's atmosphere was estimated by observing stars with
flux-calibrated spectra in a variety of observing modes (Table
\ref{tab:throughput}). The observing modes used are GPI's
standardized instrument configurations for each spectral filter:
{\it Direct} -- neither the apodizer, Lyot stop, nor the occulting
mask are in the optical path, {\it Unblocked} -- the apodizer and
the Lyot stop are in the optical path, but the occulting mask is
not, and {\it Coron} -- the apodizer, Lyot stop, and the occulting
mask are all in the optical path. The specific masks and stops used
depend on the chosen filter. The data were reduced using the GPI
DRP, and were typically sky subtracted using sky frames obtained
immediately before or after the observations of the calibration
target. The flux ($F_{\lambda}$) as a function of wavelength
($\lambda$) for each filter was measured by either summing the
counts (in ADU/coadd) within each wavelength slice of the reduced
sky-subtracted data cubes for open-loop observations, or by
performing aperture photometry for closed-loop observations. For HD
8049 B, the aperture was set at three times the measured FWHM, to
miminize the effect of scattered light from the primary star. The
remaining targets with closed-loop observations used a larger
aperture of nine times the FWHM. As the calibration targets were
observed at a relatively low airmass ($< 1.5$), the
airmass-dependent extinction was estimated to be $\lesssim 10\%$
(corresponding to a change in throughput of $\lesssim 1\%$) based on
literature empirical relations \cite{Tokunaga}. For each target the
measured flux was then compared to the flux-calibrated spectra in
the literature ($M_{\lambda}$, reference given in Table
\ref{tab:throughput}), convolved to the resolution of the GPI
measurements and corrected for the airmass-dependent extinction, to
determine the total system throughput ($T_{\lambda}$) as a function
of wavelength.
%equation for throughput
\begin{equation}
\label{eq:throughput}
T_{\lambda}=\left(\frac{F_{\lambda}}{M_{\lambda}}\right)\left(\frac{ghc}{At
\lambda d\lambda}\right)
\end{equation}
Equation \ref{eq:throughput} shows the expression used in the
throughput calculation, where $g=3.04$~e${^-}$/ADU is the gain of
the detector, $h$ is the Planck constant, $c$ is the speed of light,
$A=46.05$~m$^2$ is the collecting area of the Gemini South
telescope, and $t$ is the exposure time in seconds/coadd.

\begin{figure}[h]
\begin{center}
\includegraphics[width=15cm,natwidth=576,natheight=353]{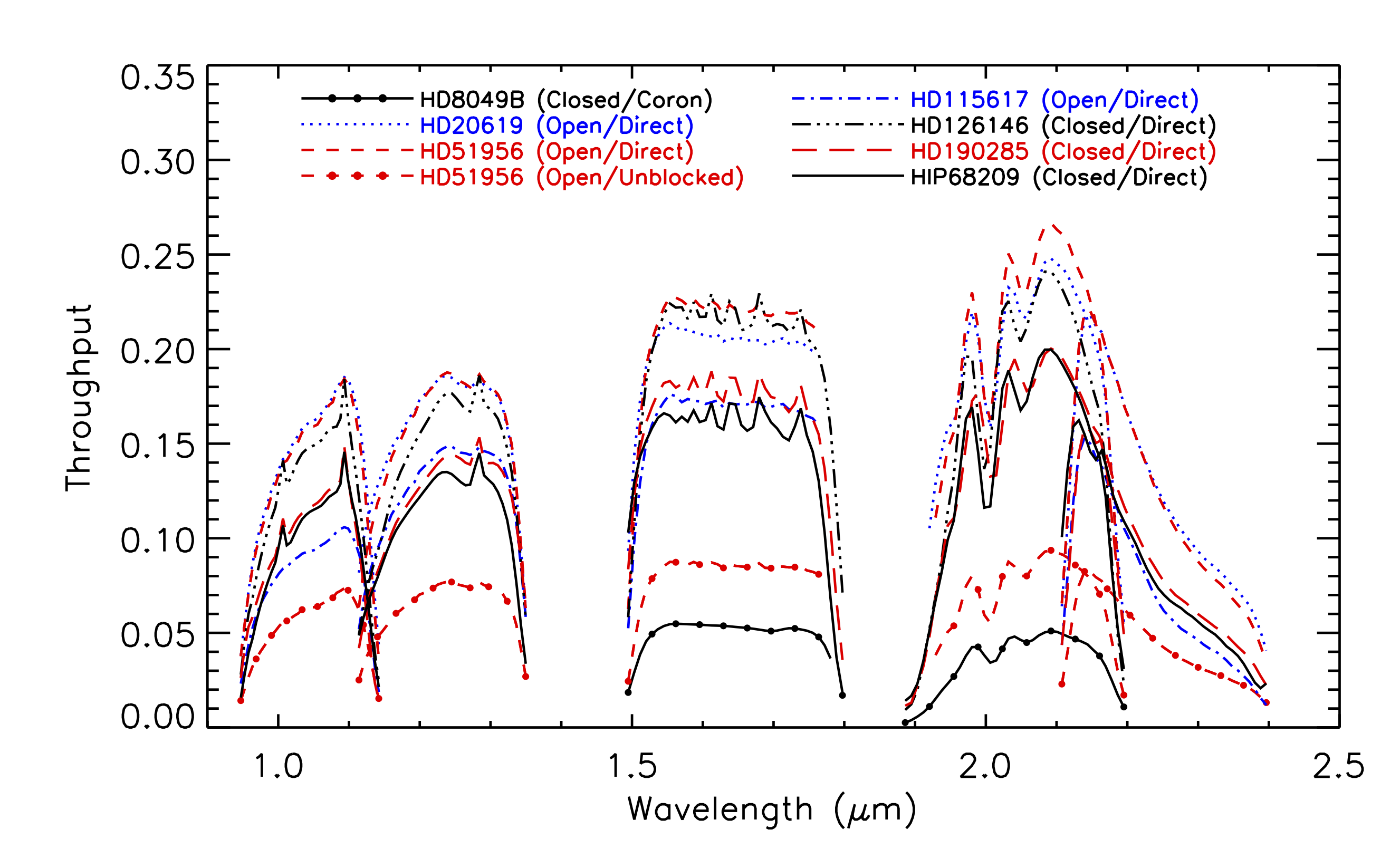}
\end{center}
\caption{The throughput of the GPI instrument, the Gemini telescope
and the atmosphere of the Earth measured using seven calibration
targets observed during the GPI commissioning runs (see Table
\ref{tab:throughput}). The different targets are differentiated by
linestyle, with filled circle symbols denoting observations taken in
either {\it Unblocked} mode (HD 51956) or {\it Coron} mode (for HD
8049 B).} \label{fig:throughput}
\end{figure}

The resulting throughput curves for each target are plotted in
Figure \ref{fig:throughput}, differentiated by object and
observation mode. GPI achieves between $\sim$ 10--25\% total system
throughput in {\it Direct} observing modes without the coronagraphic
optics, and around 4--9\% total system throughput in coronagraphic
modes. The significant drop in the throughput between the
observations taken in the {\it Direct} and the {\it Unblocked} or
{\it Coron} modes is caused by the presence of the apodizer, and to
a lesser extent the Lyot stop within the optical path. This is by
design and as expected based on the coronagraph optics properties.
We caution that there is still significant scatter in the measured
throughput when comparing different sources, which may be due to
some combination of uncertain absolute flux calibrations for these
targets, variations in atmospheric transparency, and observational
uncertainties. The throughput inferred from the HD 51956 {\it
Unblocked} $H$-band spectrum is unfortunately only about half of the
design goal for GPI total system throughput. However, GPI
observations are typically expected to be contrast limited rather
than photon limited so the impact of this on science observations
may not be substantial.\footnote{Separate measurements of throughput
to the AO system and wavefront sensor detectors (not presented here)
show that throughput to those detectors is in line with
expectations. A lower than expected throughput in the IFS subsystem
does not impact the performance of the GPI AO system in any way.}
The throughput measured using the {\it Coron} observations of HD
8049 B may be underestimated due to the small aperture size used.
The significant decrease in throughput measured beyond
$\sim$2.2~$\mu$m in the $K2$ bandpass is currently under
investigation.

\begin{table}[h!]
\centering \caption{Average throughput measurements for {\it Direct}
observations at central wavelength of each filter}
\begin{tabular}{ccc}
\hline
 Filter & Central Wavelength & Throughput \\
        &     ($\mu$m)       & (without coronagraph) \\
\hline \hline
$Y$ & 1.05 & $0.137\pm0.027$\\
$J$ & 1.24 & $0.166\pm0.023$\\
$H$ & 1.65 & $0.195\pm0.025$\\
$K1$ & 2.05 & $0.206\pm0.025$\\
$K2$ & 2.26 & $0.089\pm0.025$\\
\hline
\end{tabular}
\label{tab:throughput_values}
\end{table}

The average throughput measured at the central wavelength of each
filter for the observations taken in the {\it Direct} mode is given
in Table \ref{tab:throughput_values}. While consistent throughput
measurements are seen for targets observed on the same night (e.g.
HD 20619/HD 51956) and during the same run (e.g. HD 190285/HIP
68209), there is a $\sim$30\% variation in the measured throughput
over the six-month period for which data are available. The cause of
this variation is uncertain. The observations were all obtained
under good atmospheric conditions, although gray extinction from
light cloud cover above the observatory could induce a significant
change in the measured throughput at certain wavelengths.

\section{Ongoing development of datacube assembly algorithms} \label{sec:ongoing}
    % Upcoming improvements ?
    % Ongoing work
    % Improvements under development
    % Algorithms currently under development
    % New Datacube extraction development
    % Upcoming improvements to datacube assembly
    % Ongoing development of datacube assembly algorithms ?   - i think the best so far... but not great
    % Ongoing development of datacube extraction method ?? - ya...    % datacube construction? - datacube assembly! I like it - plus our primitives are "assemble spectral datacube" Right

%    \todo{I'm not sure the first sentence is appropriate. Does this algorithm use the PSFs to improve the spectral resolution? I thought it just improved the data quality.. }

\begin{figure}[h]%[H]
  \begin{center}
  \begin{tabular}{c}
  \includegraphics[width=7cm,natwidth=393,natheight=385]{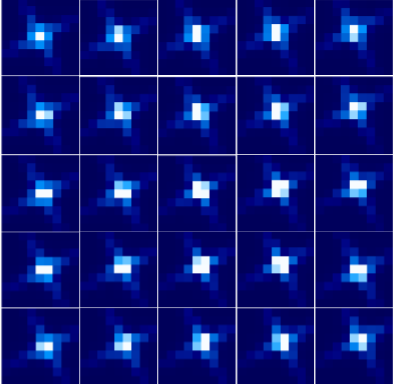}
    \includegraphics[width=7cm,height=7cm,natwidth=393,natheight=385]{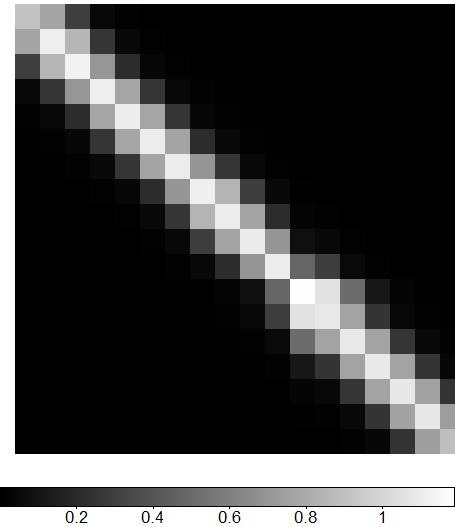}
  \end{tabular}
  \end{center}
  \caption{Left: The same simulated micro-lens PSF at $1.5\mu m$ as seen on the detector, sampled differently due to 25 different
sub-pixel center locations. Just the center of the 64$\times$64 PSF
is represented. Right: Example of the $(\bold{P}'\bold{P})$ matrix
to be inverted. The system is well conditioned as the matrix is
almost diagonal.}
%>>>> use \label inside caption to get Fig. number with \ref{}
  \label{fig:mlenspsf}
  \end{figure}

The individual lenslet PSFs are not Nyquist sampled by the detector,
having FWHM between $\sim$$1.5-1.8$ pixels \cite{Ingraham2this}. It
is possible to slightly improve the spectral resolving power by
taking into account of the micro-lens PSF extent in the datacube
extraction process. We report here a second GPI DRP primitive for
datacube extraction, ``Assemble Spectral Datacube using mlens PSF",
that uses a model of the micro-lens PSF to disentangle the flux
variations along each spectrum. This primitive has been implemented
in the GPI DRP since November 2012 but is undergoing improvements.
The primitive utilizes the GPI Data Simulation Tool\cite{Maire2010}
to construct a set of micro-lens PSFs calculated by taking into
account the size and shape of the individual micro-lens and the
modulation transfer function (MTF) of the detector pixels.
% \todo{The obvious next step here is to replace the DST simulated PSFs with the actual measured ones from Patrick's work. Should at least mention this here I think.JM: yeah this is mentioned later in the text}
The PSFs are sampled at the spatial resolution of the detector, and
calculated for each spectral channel and for each PSF center
sub-pixel location. Figure \ref{fig:mlenspsf} (left) illustrates how
the simulated 1.5$\mu m$ micro-lens PSF changes as the central
location is shifted by a fifth of a pixel in each direction.
% \todo{I'm not sure this figure is very relevant because in practice the observed PSFs don't look much like these models. The diffraction from the square is much less evident, there's more geometrical aberrations etc.JM: this is no more relevant since we have the ePSF, but ePSF wasn't existing few weeks ago...but yeah we mention that the ePSF will be implemented}
% The 64$\times$64 micro-lens PSF  is determined for each sub-pixel location of the PSF center, at a fifth of a pixel in both x and y direction, giving 25 micro-lens  PSFs for each waveband of the 37 spectral channels, giving thus a cube of 925 64$\times$64 microlens PSFs for each spectral band.

 %-------------
%-------------

 The intensity $I$ of a detector pixel at position ($x$,$y$) in the image can be considered as the sum over all $N$ spectral channels  $\lambda$ of the product of the micro-lens PSF $P (x-x_c(\lambda), y-y_c(\lambda), \lambda) $ centered on $(x_c(\lambda),y_c(\lambda))$ by the flux $F(\lambda)$ of the target in this channel, such as
 \begin{equation}
    \label{eq1}
    I(x, y)=\sum\limits_{\lambda=1}^{\lambda=N} F(\lambda) \times P \left(x-x_c(\lambda), y-y_c(\lambda), \lambda \right).
   \end{equation}
This technique aims to recover the flux of each individual spectrum
while mitigating the  contamination by adjacent spectral channels. A
term for taking into account correlated detector noise can be added
to study noise propagation, but here we treat the term as negligible
since this extraction happens after the removal of correlated noise
in the raw images by the pipeline.

Using Equation \ref{eq1}, it is then possible to form a system of
equations using $M$ pixels of a specific spectrum with intensities
$I_m(x_m,y_m)$ where ($m=1$,$\ldots$,$M$). Letting
$x_m'=x_m-x_c(\lambda)$ and $y_m'=y_m-y_c(\lambda)$, this system can
be expressed in a matrix form $\bold{I}=\bold{P}\cdot\bold{F}$ as
  \begin{equation}  \label{eq2}
  \left( \begin{array}{c}
                    I_1 \\
                    I_2 \\
                    \vdots\\
                    I_M \end{array} \right)
                    =
  \left( \begin{array}{cccc}
P(x_1',y_1', \lambda_1) & P(x_1',y_1', \lambda_2) & \ldots & P(x_1',y_1', \lambda_N) \\
P(x_2',y_2', \lambda_1) & P(x_2',y_2', \lambda_2) & \ldots & P(x_2',y_2', \lambda_N) \\
                \vdots & \vdots & \ddots & \vdots \\
  P(x_M',y_M', \lambda_1) & P(x_M',y_M', \lambda_2) & \ldots & P(x_M',y_M', \lambda_N)
                \end{array} \right)
                \left( \begin{array}{c}
                    F(\lambda_1) \\
                     F(\lambda_2) \\
                    \vdots\\
                     F(\lambda_N) \end{array} \right).
   \end{equation}
Solving this for $\bold{F}$ gives the standard unweighted
least-squares solution\cite{Cowan1998}
   \begin{equation}
\bold{F}= (\bold{P}'\bold{P})^{-1}\bold{P}'\bold{I}.
   \end{equation}

As $\bold P$ is a $M\times N$ matrix, this requires in particular
that $N\le M$, that is, the number of unknown fluxes or spectral
channels is smaller than or equal to the number of detector pixels
of the spectrum considered to resolve the system. The system is well
conditioned; the condition number associated with this linear
equation is very low since $(\bold{P}'\bold{P})$ is almost diagonal.
An example of the matrix $(\bold{P}'\bold{P})$ is represented in
Figure \ref{fig:mlenspsf} (right),  showing that this algorithm is
appropriate to resolve the system. This inversion technique is
applied on each spectrum individually. Different choices of pixel
configuration have been tested, mainly different number of columns
of pixels along the spectrum. As GPI spectra are about 18 pixels
long in $H$-band, 1, 3 or 5 columns of 18 pixels can be selected to
form the observed $\bold{I}$ vector. These columns of pixels are
tilted for the spectra at the edges of the field-of-view, according
to the corresponding wavelength solution.

\begin{figure}
  \begin{center}
  \begin{tabular}{c}
    \includegraphics[width=12cm,natwidth=393,natheight=385]{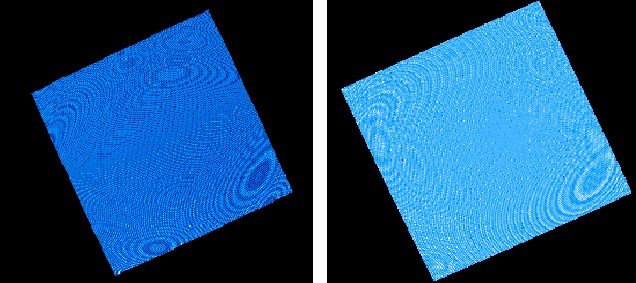}\\
  \includegraphics[width=12cm,natwidth=393,natheight=385]{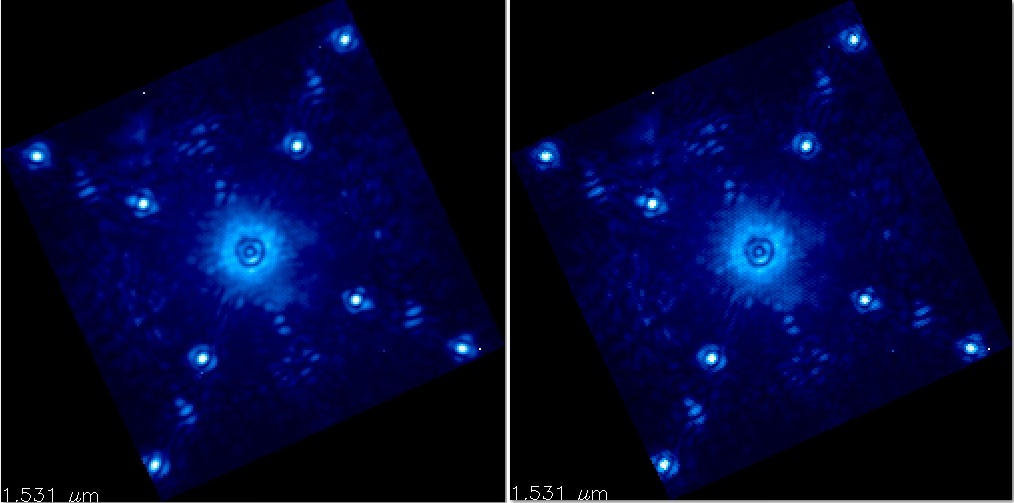}\\
  \includegraphics[width=12cm,natwidth=393,natheight=385]{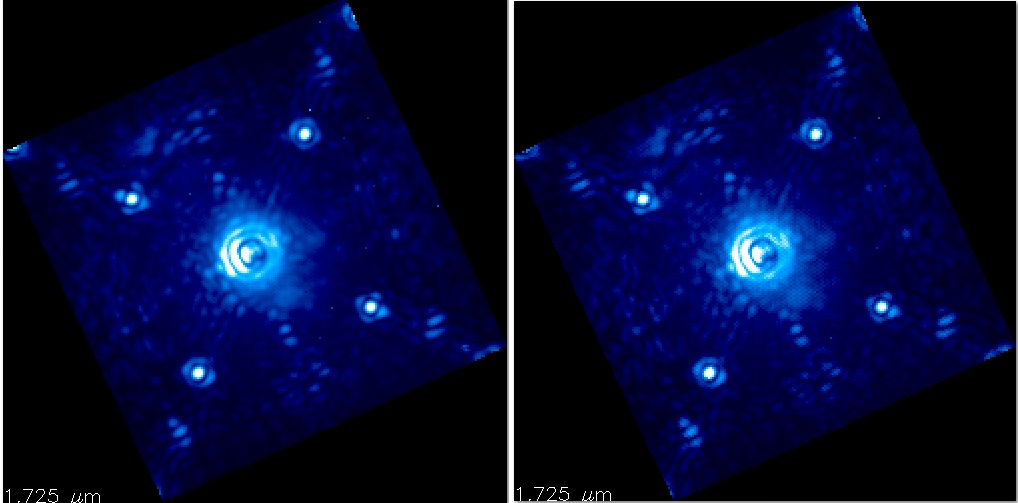}
  \end{tabular}
  \end{center}
  \caption{ Comparison of images obtained by integrating the signal over a rectangular aperture along the spectra (left) and by using an inverse method taking into account the simulated micro-lens PSF (right) for GPI data taken during UCSC Integration \& Test period (middle and bottom) and for simulated flat-field lamp (top).
%   \todo{The figures look less good on the right with the inversion method. More vertical striping visible in the data. Why is this? This figure doesn't really help make the case that the inversion method is better. The top panel for simulated flat field data is not that useful to show either. What is a reader supposed to conclude from looking that this?JM:this is just to show that the method is actually working to extract datacube. i agree, we can have a better one, for instance by making a comparison of extracted spectrum when there is just one emisson line. Could probably work on that in the next week. }
  }
%>>>> use \label inside caption to get Fig. number with \ref{}
  \label{fig:UCSC}
  \end{figure}

The number of spectral channels is chosen to be slightly larger than
the number of channels defined by the detector-sampling-limited
effective spectral resolution, i.e 10 channels given the
$\sim$1.5~pixel FWHM of the micro-lens PSF on an 18~pixel-length
spectrum. We performed tests with 9, 12, 15 and 18 spectral channels
using simulated flat-field images and GPI data taken during the
integration and testing of the instrument (Figure \ref{fig:UCSC}).
The flat-field images were simulated using a flat spectrum with same
intensity for all micro-lenses, then datacubes were extracted using
the two methods. We then measured the ratio of the standard
deviation of pixel intensities in a slice of the datacubes to the
mean intensity in the same slices. These ratios, 4.3\% for the
3$\times$1 pixel box method and 2.9\% for the inversion method,
shows that the inversion method was able to extract the flux more
accurately in this test. The Moir\'e pattern visible in the flat-field
image (Figure \ref{fig:UCSC}, top) has been identified to come from
a systematic bias in the determination of the wavelength solution,
for which a new method of measurement has been
developed\cite{Schuylerthis}. The inversion method method can
extract datacube with high fidelity, but it is more sensitive to any
bias in the wavelength solution or uncorrected shifts due to flexure
effect than the integration method, giving a more pronounced
checkerboard pattern, as shown in Figure \ref{fig:UCSC} (middle,
bottom).
 %-------------
%-------------

The performance of this inversion technique is dependent upon an
accurate model of the lenslet PSF. It has been demonstrated using
GPI calibration data that the micro-lens PSF shape varies across the
field of view\cite{Ingraham2this}, presenting some specific
elongations for micro-lenses towards the edges of the field-of-view.
This semi-empirical model of the micro-lens PSF will be implemented
in a future version of the inversion technique for datacube
extraction. Updated wavelength solutions\cite{Schuylerthis} should
also provide a significant improvement to the inversion techniques
datacube output quality. This inversion method has not yet been
employed much on on-sky observational data, pending these
improvements. A similar technique with adaptive fitting of the
flexure effects is also under development\cite{Zackthis}.
%\todo{BM - we should cite Zack's work here too}
% \todo{We need to explicitly mention also Zack's paper as another alternative version of this with some similar aspects and some differences. Would be useful to mention some of the pros and cons of each?JM: Zach paper is mentioned previously in the text. Yes it would be nice that Zach makes a comparison of his method with this one that have been released since 2012 and presented in GPI meeting since 2008.... }

\section{Conclusion \& Future work} \label{sec:conc}
%\todo{Ok, this section needs work}

    We have seen how to perform a photometry calibration  to establish a relation between observed instrumental counts and expected amount of photons. On base of this calibration, all observed objects and the  image itself can be transformed from counts to physical units.

 We measured the throughput expressed in all filter wavebands that include  the various optical components of GPI, the telescope optics
as well as the transmission characteristics of the atmosphere, by
using a set of known calibration stars.  Throughput variability will
be characterized in more details as further observations proceed.

 We have seen that the pipeline can deliver high-fidelity spectrum of observed objects and provides uncertainties on this measurement. Improving the accuracy of the photometry requires to better understand all systematic errors and to optimize each step of the reduction. It could be envisaged to improve the pipeline by implementing error propagation through each step of the reduction to produce a cube giving uncertainty on each spaxel of a datacube.

Datacube extraction itself is crucial to improve the accuracy of the
photometry for which we proposed a new datacube extraction method
that can distangle the flux variations along each spectrum and to
mitigate spectral cross-talks. Further improvements by use of
empirical micro-lens PSF will be implemented.

\section{Acknowledgments}
The Gemini Observatory is operated by the Association of
Universities for Research in Astronomy, Inc., under a cooperative
agreement with the NSF on behalf of the Gemini partnership: the
National Science Foundation (United States), the National Research
Council (Canada), CONICYT (Chile), the Australian Research Council
(Australia), Minist\'erio da Ci\'encia, Tecnologia e Inova\c{c}\=ao
(Brazil), and Ministerio de Ciencia, Tecnolog\'ia e Innovaci\'on
Productiva (Argentina). The Dunlap Institute is funded through an
endowment established by the David Dunlap family and the University
of Toronto.

\bibliography{biblioGPIspie2014}   %>>>> bibliography data in report.bib
\bibliographystyle{spiebib}   %>>>> makes bibtex use spiebib.bst

\end{document}